\documentstyle[aps,axodraw,preprint,prd]{revtex}

\def\bfeps{\mbox{\boldmath$\epsilon$}}
\def\bfsig{\mbox{\boldmath$\sigma$}}
\def\bp{{\bf p}}
\def\bk{{\bf k}}
\def\H2{H^{**}}
\def \to {\rightarrow}

\begin{document}
\preprint{\ \vbox{ \halign{&##\hfil\cr
       AS-ITP-2001-0014\cr\cr}} } \vfil
\draft
\title{Revisiting spin alignment of heavy meson in its inclusive production}
\author{J.P. Ma}
\address{Institute of Theoretical Physics,\\
Academia Silica, \\
P.O.Box 2735, Beijing 100080, China\\
e-mail: majp@itp.ac.cn} \maketitle

\begin{abstract}
\ In the heavy quark limit inclusive production rate of a heavy
meson can be factorized, in which the nonperturbative effect
related to the heavy meson can be characterized by matrix
elements defined in the heavy quark effective theory. Using
this factorization, predictions for the full spin density matrix
of a spin-1 and spin-2 meson can be obtained and they are
characterized only by one coefficient representing
the nonperturbative effect. Predictions for spin-1 heavy
meson are compared with experiment performed at $e^+e^-$
colliders in the energy range from $\sqrt{s}=10.5$GeV
to $\sqrt{s}=91$GeV, a complete agreement is found for
$D^*$- and $B^*$-meson. For $D^{**}$ meson, our
prediction suffers a large correction, as indicated
by experimental data. There exists another approach
by taking
heavy mesons as bound systems, in which the
total angular momentum of the light degrees of freedom
is $\frac{1}{2}$ and $\frac{3}{2}$ for spin-1 and spin-2 meson
respectively, then the diagonal parts of spin
density matrices can be obtained.
However, there are distinct differences in the predictions
from the two approaches and they are discussed in detail.

\end{abstract}

\vfill\eject\pagestyle{plain}\setcounter{page}{1}

\noindent
{\bf 1. Introduction}
\par\vskip20pt
In the heavy quark limit the momentum of a heavy meson containing
one heavy quark can be approximated as the momentum of the heavy
quark. This fact allows us to study properties of a heavy meson
by starting from QCD directly. Activities of the study have leaded
to the birth
of the heavy quark effective theory(HQET)\cite{HQET,Review},
with this effective theory a consistent and systematic expansion
in the inverse of the heavy quark mass can be preformed.
HQET has been widely used in studies of decays of heavy hadrons,
in this work we apply HQET for inclusive productions of
a polarized heavy meson, where the heavy meson is a spin-1 or
spin-2 particle.
\par
In its inclusive production a heavy meson is formed with a heavy
quark $Q$ and with other light degrees of freedom, the light degrees
can be a system of light quarks and gluons. Because its large mass
$m_Q$ the heavy quark is produced by
interactions at short distance. Therefore the production can be
studied with perturbative QCD. The heavy quark, once produced, will
combine light degrees of freedom to form a hadron, the formation
is a long-distance process, in which
momentum transfers are small, hence the formed
hadron will carry the most momentum of the heavy quark. The
above discussion implies the production rate can be
factorized, in which the perturbative part is for
the production of a heavy quark, while the nonperturbative
part is for the formation. For the nonperturbative
part a expansion in the inverse of $m_Q$ can systematically
be performed in the framework of HQET. This type
of the factorization was firstly used in parton fragmentation
into a heavy hadron\cite{Ma}.
\par
In this work we will use the factorization to predict
the spin density matrix of a spin-1- or spin-2 heavy meson
in its inclusive production. It turns out that the matrix
is determined by perturbative coefficients and  by one
nonperturbative parameter, which is a ratio of matrix elements
defined in HQET. The predictions are compared with experiment
performed at $e^+e^-$
colliers in the energy range from $\sqrt{s}=10.5$GeV
to $\sqrt{s}=91$GeV, a good agreement is found for
$D^*$- and $B^*$-meson.
\par
The diagonal part of spin density matrices have been studied
before\cite{FP}, in which the total angular momentum $j$ of
the light degrees of freedom in the heavy meson is taken as
$\frac{1}{2}$ and $\frac{3}{2}$ for the spin-1- and spin-2 meson,
respectively, i.e., $j$ takes the smallest of possible values.
It should be noted that this property of
a heavy meson has been not derived from QCD, because solutions
of eigenvalue problem of QCD hamiltonian in the heavy quark
limit are not known. This property may be argued by that
the system of light degrees of freedom with larger value of $j$
is produced with a suppressed probability. For $B^*$ this property is
experimentally supported by observing the ratio of the
production rates of spin-0-and spin-1 states, while
for charm mesons the ratio has a $20\%$ deviation
from the prediction based on the property.
In our approach we will not take the property into
account. In other word our results are derived on a
more general ground and corrections from higher orders
in $m_Q^{-1}$ can be systematically added. It is
interesting to note that the obtained predictions
of the spin density matrix of $D^*$ also agree with
experiment well, in which one expects large corrections
form higher orders of $m_c^{-1}$. For a spin-2 heavy meson
a parameter is introduced for its production in the approach
in \cite{FP}, and
this is the only parameter characterizing
the angular distribution of its decay. In our approach
the corresponding distribution is predicted as an isotropic
one. Experimentally this distribution is measured
with a limited data sample for $D^{**}$. Although the distribution
is not well determined in experiment, the measured
distribution does not agree with our prediction, indicating
that the correction to our prediction made in the heavy
quark limit is substantial, this is also theoretically expected,
because the binding energy of $D^{**}$ is quite large.
We will discuss
in detail the difference between the two approaches.
\par
In this work we consider the inclusive production at $e^+e^-$
colliders, where the initial beams are unpolarized. Our results
can be generalized to the case with polarized beams, and also to
inclusive production at hadron colliders or at $ep$-colliders. It
is straightforward to obtain predictions for other inclusive
predictions. Our work is organized as the following: In Sect.2 we
give our result for spin-1 meson and compare our predictions with
experiment. The differences between our approach and that in
\cite{FP} are discussed in detail. In Sect.3 we present results
for spin-2 meson. Sect.4 is our summary. Throughout this work we
take nonrelativistic normalization for heavy quarks and for heavy
mesons.
\par\vskip20pt
\noindent
{\bf 2. The density matrix for spin-1 heavy meson}
\par\vskip20pt
We denote a spin-1 heavy meson as $H^*$, which contains one
heavy quark $Q$. We study the inclusive production
\begin{equation}
e^+({\bf p})+e^{-}(-{\bf p})\to H^*({\bf k}) +X,
\end{equation}
where the three momenta are given in the brackets. In the
process we assume that the initial beams are unpolarized.
We denote the helicity of $H^*$ as $\lambda$ and
$\lambda=-1,0,1$. All information about the polarization
of $H^*$ is contained in a spin density matrix, which may
be unnormalized or normalized, we will call them
unnormalized or normalized spin density matrix, respectively.
The unnormalized spin density matrix can be defined as
\begin{equation}
R(\lambda, \lambda',{\bf p},{\bf k})
 = \sum_X \langle H^*(\lambda)X\vert {\cal T}\vert e^+e^-\rangle
         \cdot \langle H^*(\lambda')X\vert {\cal T}\vert e^+e^-\rangle^*,
\end{equation}
where the conservation of the total energy-momentum and the spin
average of the initial state is implied. ${\cal T}$ is the
transition operator. The cross-section with a given helicity
$\lambda$ is given by:
\begin{equation}
\sigma(\lambda) = \frac{1}{2s} \int \frac {d^3k}{(2\pi)^3}
R(\lambda, \lambda,{\bf p},{\bf k}).
\end{equation}
From Eq.(2) the normalized spin density matrix is defined by
\begin{equation}
\rho_{\lambda\lambda'}({\bf p},{\bf k}) =
 \frac {R(\lambda, \lambda',{\bf p},{\bf k})}
    {\sum_\lambda R(\lambda, \lambda,{\bf p},{\bf k})}.
\end{equation}
It should be noted that the normalized spin density matrix
is measured in experiment.
\par
The contribution to $R(\lambda, \lambda',{\bf p},{\bf k})$ can be
characterized by Feynman diagrams. With the factorization
discussed in the introduction, each diagram can be divided into
two parts. For the process in Eq.(1) the contribution at
tree-level is given in Fig.1, where we divide the diagram with a
horizontal broken line into two parts. The upper part with the
black box represents the formation, while the lower part
represents the inclusive production of a heavy quark $Q$, which is
not necessarily on-shell here. At loop-level it is possible that
there are gluon exchanges between quark lines and between the
lower part and the black box. The effect of the exchanges between
the lower part and the black box is at higher orders in
$m_Q^{-1}$, we will neglect it in this work. The exchanges of
gluons between quark lines will in general bring some infrared
singularities in the lower part. We can show that at the leading
order of $m_Q^{-1}$ the heavy quark lines connecting the lower-
and upper part represent on-shell quarks. Hence, the lower part
with possible gluon exchanges is responsible for inclusive
production of an on-shell quark $Q$, and it is free from infrared
singularities.
\par
\begin{center}
\begin{picture}(400,400)(0,0)
\SetWidth{2.0}

\DashLine(50,0)(100,0){10}
\Text(75,10)[]{$\gamma, ~Z$}

\ArrowLine(300,0)(100,0)

\DashLine(300,0)(350,0){10}
\Text(325,10)[]{$\gamma, ~Z$}

\ArrowLine(100,0)(100,200)
\Text(80,100)[]{{\Large q}}

\Line(300,0)(300,200)
\GBox(100,200)(300,250){0.5}

\DashLine(200,-25)(200,275){5}
\DashLine(75,160)(325,160){5}

\end{picture}
\end{center}
\vskip 10mm
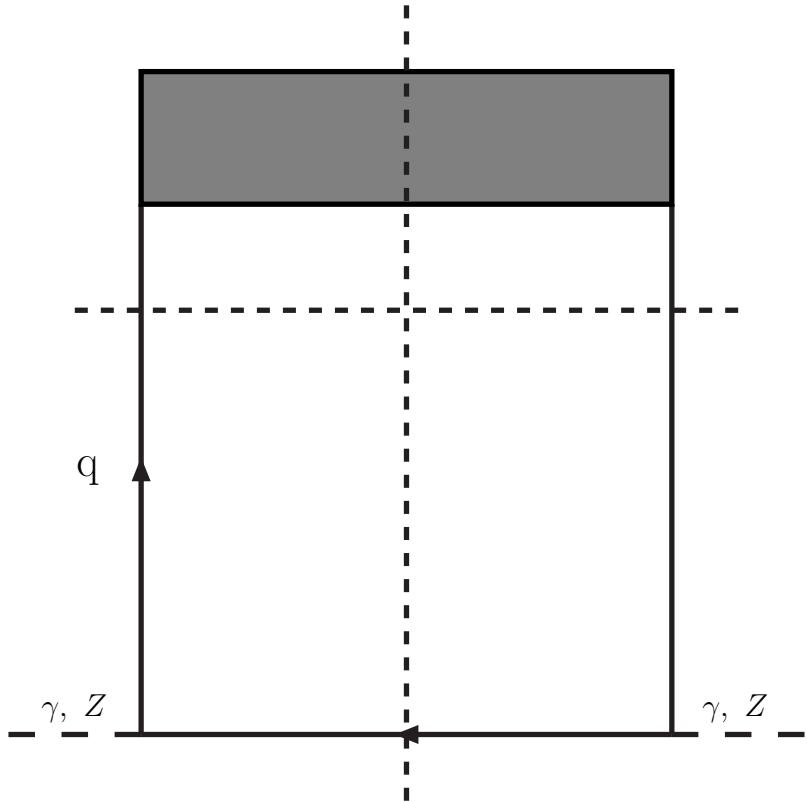
{\bf Fig.1}: The diagram for the contribution at tree-level. The line is for
heavy quarks. The vertical broken line is the Cutkosky cut.
\par\vskip20pt
With the diagram the density matrix can be written as
\begin{equation}
R(\lambda, \lambda',{\bf p},{\bf k}) =
\int \frac {d^4q}{(2\pi)^4}
    T_{ji} (q) \Gamma_{ij}(\lambda,\lambda',q,k),
\end{equation}
where $T_{ji} (q)$ is for the lower part and can be calculated
with perturbative theory, $\Gamma_{ij}(\lambda,\lambda',q,k)$
represents the formation of the heavy meson, which is defined
as:
\begin{equation}
\Gamma_{ij}(\lambda,\lambda',q,k)
 =\int d^4x e^{-ix\cdot q} \sum_X
    \langle 0\vert Q_i(0) \vert H^*(\lambda')X\rangle
    \langle X H^*(\lambda)\vert \bar Q_j(x)\vert 0 \rangle.
\end{equation}
The indices $i,j$ stand for Dirac- and color indices, they connect the
lower- and the upper part through the vertical quark lines. $Q(x)$
is the Dirac field for the heavy quark $Q$. In the heavy quark
limit the heavy meson carries the most momentum of the heavy quark,
the difference between the momentum of the heavy quark and
that of the heavy meson is at order of $\Lambda_{QCD}$.
This scale $\Lambda_{QCD}$ is much smaller than $m_Q$ and $\sqrt s$,
which are scales appearing in the lower part of Fig.1.
This fact
indicates that we can expand $\Gamma_{ij}(\lambda,\lambda',q,k)$
in $\Lambda_{QCD}$.
With this expansion for
$\Gamma_{ij}(\lambda,\lambda',q,k)$ it results in that the density
matrix given in Eq.(6) is expanded in $m_Q^{-1}$. For doing this
we use HQET, in which the Dirac field is expanded in the field
of HQET:
\begin{eqnarray}
Q(x) &=&e^{-im_Qv\cdot x}\left\{ 1+\frac 1{2m_Q}i\gamma \cdot D_T\right\}
h_v(x)+{\cal O}(\frac 1{m_Q^2})  \nonumber \\
&&+({\rm terms\ for\ anti-quark}),
\end{eqnarray}
where $v$ is the velocity of $H^*$, $h_v(x)$ is the heavy quark field in
HQET, $D_T^\mu=D^\mu-v\cdot D v^\mu$ and $D^\mu$ is
the covariant derivative. Substituting this into $\Gamma_{ij}(\lambda,\lambda',q,k)$,
the integrand depends on $x$ through $\bar h_v(x)$
and through a exponential factor,
the $x$-dependence through $\bar h_v(x)$ is controlled by the scale at order of
$\Lambda_{QCD}$, at leading order this dependence can be neglected.
We then have:
\begin{equation}
\Gamma_{ij}(\lambda,\lambda',q,k) = (2\pi)^4\delta^4(q-m_Qv)
   \sum_X \langle 0\vert (h_v(0))_i \vert H^*(\lambda')X \rangle
          \langle X H^*(\lambda)\vert (\bar h_v(0))_j \vert 0\rangle
  +\cdots ,
\end{equation}
where the $\cdots$ stand for neglected terms, whose contributions to the density
matrix are suppressed by powers of $m_Q^{-1}$. With a detailed examination
one can show these contributions are of order of $m_Q^{-2}$ or of higher orders.
At the leading order
the momentum of $H^*$ is approximated by $m_Qv$.
To analyze the structure
of $\Gamma_{ij}$ in the above equation it is convenient to work with
the rest frame of $H^*$, which is obtained from the moving frame only
by a Lorentz boost. We choose the $z$-direction to be
the moving direction of $H^*$ and denote the polarization vector of $H^*$
in the rest-frame as $\bfeps(\lambda)$. In this frame we
can define a creation operator for $H^*$:
\begin{equation}
 \vert H^*(\lambda)\rangle =a^\dagger(\lambda) \vert 0\rangle=
  \bfeps(\lambda)\cdot{\bf a}^\dagger \vert 0 \rangle .
\end{equation}
In the rest frame the field $h_v$ has two non zero components and
can be written as:
\begin{equation}
h_v(x)=\left(\begin{array}{c} \psi(x) \\ 0 \end{array}\right).
\end{equation}
With this notation the matrix in Eq.(8) takes the form:
\begin{eqnarray}
&\sum_X&  \langle 0\vert (h_v(0))_i \vert H^*(\lambda')X \rangle
          \langle X H^*(\lambda)\vert (\bar h_v(0))_j \vert 0\rangle
          \nonumber\\
&=& \left(
   \frac {1+\gamma\cdot v}{2}\cdot
    \left( \begin{array}{cc}
\langle 0\vert \psi a^\dagger (\lambda') a(\lambda) \psi^\dagger \vert 0\rangle, & 0 \\
      0, & 0 \\
    \end{array}
     \right)\cdot\frac{1+\gamma\cdot v}{2}
     \right)_{ij}.
\end{eqnarray}
The matrix element $
\langle 0\vert \psi a^\dagger (\lambda') a(\lambda) \psi^\dagger \vert 0\rangle$
is diagonal in color space, the matrix structure in spin space
can be decomposed into the unit matrix and the Pauli matrices
$\bfsig$. For the decomposition we define two operators:
\begin{equation}
O(H^*) = \frac{1}{6}{\rm Tr} \psi a_i^\dagger a_i \psi^\dagger, \ \
O_s(H^*) = \frac{i}{12} {\rm Tr} \sigma_i \psi a^\dagger_j
   a_k \psi^\dagger \varepsilon_{ijk},
\end{equation}
where $\varepsilon_{ijk}$ is the totally antisymmetric tensor. With
these operators the matrix can be written:
\begin{equation}
\langle 0\vert \psi a^\dagger (\lambda') a(\lambda) \psi^\dagger \vert 0\rangle
 = \frac{1}{3}\langle 0 \vert O(H^*) \vert 0 \rangle \bfeps^*(\lambda)
 \cdot\bfeps(\lambda') \cdot I
  + \frac{i}{3} \bfsig\cdot [\bfeps^*(\lambda)\times\bfeps(\lambda')]
\cdot \langle 0 \vert O_s(H^*) \vert 0 \rangle,
\end{equation}
where $I$ is a $2\times 2$ unit matrix in the spin space.
It should be noted that the factor
$(1+\gamma\cdot v)/2$ in Eq.(11) is a projector and
can be written as $\sum_s u(p,s)\bar u(p,s)$,
where $u(p,s)$ is the spinor of the heavy quark $Q$ in the rest frame.
This means
that the projector projects the quark state represented by the vertical
quark lines in Fig.1 to the on-shell state. Therefore, the lower part is just
for inclusive production of a on-shell heavy quark $Q$.
The two matrix elements $\langle 0 \vert O(H^*) \vert 0 \rangle$ and
$\langle 0 \vert O_s(H^*) \vert 0 \rangle$ are universal in the
sense that they do not depend on how $Q$ is produced,
they describe the long-distance process of the formation of $H^*$.
Using Eq.(8), Eq.(11) and
Eq.(13) we
can write the result for the unnormalized spin density matrix as:
\begin{equation}
R(\lambda, \lambda',{\bf p},{\bf k}) =\frac{1}{3} a({\bf p}, {\bf k})
 \langle 0 \vert O(H^*) \vert 0 \rangle \bfeps^*(\lambda)
 \cdot\bfeps(\lambda')
  +\frac{i}{3} {\bf b}({\bf p},{\bf k})\cdot [\bfeps^*(\lambda)\times\bfeps(\lambda')]
\cdot \langle 0 \vert O_s(H^*) \vert 0 \rangle.
\end{equation}
The quantities $a({\bf p}, {\bf k})$ and ${\bf b}({\bf p},{\bf k})$ characterize
the spin density matrix of the heavy quark $Q$ produced in the inclusive process:
\begin{equation}
e^+({\bf p})+e^{-}(-{\bf p})\to Q({\bf k},{\bf s}) +X
\end{equation}
where ${\bf s}$ is the spin vector of $Q$ in its rest frame and
the rest frame is related to the moving frame only by a Lorentz
boost. The unnormalized spin density matrix $R_Q({\bf s},{\bf
p},{\bf k})$ of $Q$ can be defined by replacing $H^*(\lambda)$
with $Q({\bf k},{\bf s})$ in Eq.(2). This matrix can be calculated
with perturbative theory because of the heavy mass. The result in
general takes the form
\begin{equation}
R_Q({\bf s},{\bf p},{\bf k}) =a({\bf p},{\bf k}) +{\bf b}({\bf
p},{\bf k}) \cdot {\bf s}
\end{equation}
where $a({\bf p},{\bf k})$ and ${\bf b}({\bf p},{\bf k})$ are the
same in Eq.(14). The results at tree-level for $a({\bf p},{\bf
k})$ and ${\bf b}({\bf p},{\bf k})$ may be found in \cite{BeMa}.
With Eq.(14) we obtain the normalized spin density matrix:
\begin{equation}
\rho({\bf p},{\bf k}) =\frac{1}{3}\left(
\begin{array}{ccc}
  1+P_3, & -P_+, & 0 \\
   -P_-, & 1, & -P_+ \\
  0, &-P_-,  &1-P_3
\end{array}\right),
\end{equation}
with
\begin{equation}
P_3= \frac{b_3(\bp,\bk)}{a(\bp,\bk)}
            \cdot \frac{\langle 0 \vert O_s(H^*) \vert 0 \rangle}
             {\langle 0 \vert O(H^*) \vert 0 \rangle},\ \ \
P_\pm = \frac{b_1(\bp,\bk)\pm ib_2(\bp,\bk)}{\sqrt{2} a(\bp,\bk)}
           \cdot \frac{\langle 0 \vert O_s(H^*) \vert 0 \rangle}
             {\langle 0 \vert O(H^*) \vert 0 \rangle},\ \ \
\end{equation}
The indices of matrices in Eq.(17) run from -1 to 1.
\par
Before confronting experimental results we give the following comments to
our results:
\par
1). If one uses perturbative theory to calculate the coefficients in Eq.(14) or
Eq.(16), one will encounter terms with large logarithm $\ln(m_Q/\sqrt s)$. One needs
to resume these terms. One way to resume is to use the well known factorization
formula for the process in Eq.(1), where the heavy meson $H^*$ is produced
through parton fragmentation. The definitions of the spin-dependent
fragmentation functions can be found in\cite{Ji,SST}.
Then one applies the approach here to
the fragmentation functions as used in \cite{Ma}. These functions can be calculated
at the energy scale $\mu =m_Q$ with perturbative QCD. Through renormalization
group equations one obtains the functions and
then the coefficients at the energy scale $\mu =\sqrt s$.
Before the resummation the coefficients at tree-level are singular in the energy
of $H^*$, but the ratio $P_3$ and $P_\pm$ are regular.
After the resummation is performed, the coefficients become regular.
\par
2). If the parity is conserved and T-odd effects can be neglected,
which are proportional to $m_Q/\sqrt s$, then ${\bf b}({\bf
p},{\bf k})=0$, i.e., the heavy quark is unpolarized, it leads to
that $H^*$ is unpolarized too. We obtain in this case
$\rho_{00}=\rho_{-1-1}=\rho_{11}=1/3$, and all nondiagonal
elements of the spin density matrix are zero. It should be noted
that in general the nondiagonal elements can be nonzero even if
the parity is conserved, e.g., a light vector meson can have
tensor-polarization, which partly corresponds to the matrix
element $\rho_{-\lambda\lambda}$ for $\lambda=\pm 1$.
Unfortunately, the spin of $H^*$ is analyzed in experiment with
its parity-conserving two-body decay, where the polarization of
the decay products is not observed. This leads to that the
polarization of the heavy quark, i.e., the contributions from
${\bf b}(\bp,\bk)$ to the spin matrix element, can not be
determined in experiment, this is the so-called "no-win"
theorem\cite{FP}.
\par
3). Our approach can be used not only for the process in Eq.(1),
but also for other processes. For other inclusive productions the same
analysis can be done, where one
needs to replace the lower part in Fig.1 and $T_{ji}(q)$ in Eq.(5)
with those for the corresponding production of $Q$, while the
upper part in Fig.1 and $\Gamma_{ij}$ in Eq.(5) remain the same.
With the analysis presented here it is straightforward to obtain
predictions for other inclusive productions, and
predictions always takes the form as given in Eq.(14) or in
Eq.(17), where the coefficients $a({\bf p},{\bf k})$ and ${\bf
b}({\bf p},{\bf k})$ should be replaced by those characterizing
the spin density matrix of $Q$ in the
corresponding processes. Without knowing these coefficients we can
always conclude that in any inclusive production of $H^*$ we have
$\rho_{00}=1/3$ and $\rho_{-11}=\rho_{1-1}=0$ in the heavy quark
limit. These predictions may be tested in other processes, e.g.,
in inclusive production at electron-proton colliders or at
hadron-hadron colliders. It is to the author's knowledge that the
spin-measurements for $B^*$, $D^*$ and $D^{**}$ have been done
only at $e^+e^-$-colliders by those experimental groups, whose
results are listed in this work. The list may be incomplete. It
would be interesting to test our predictions in experiment at an
electron-proton- or a hadron-hadron collider.
\par
The experiments to measure the polarization of $B^*$ are performed
at LEP with $\sqrt s =M_Z$ by different experiments groups. To
measure the polarization the dominant decay $B^*\to \gamma B$ is used,
where the polarization of the photon is not observed. Because
the parity is conserved and the distribution of the angle
between the moving directions of $\gamma$ and of $B^*$ is measured,
one can only determine the matrix element $\rho_{00}$. If we denote
$\theta$ is the angle between the moving directions
of $B^*$ and of $\gamma$ in the $B^*$-rest frame and $\phi$
is the azimuthal angle of $\gamma$, then the angular distribution
is given by $W_{B^*\to B\gamma}(\theta,\phi)\propto \sum_{\lambda\lambda'}
\rho_{\lambda\lambda'}(\delta_{\lambda\lambda'}-Y_{1\lambda}(\theta,\phi)
Y^*_{1\lambda'}(\theta,\phi))$. Integrating over $\phi$ and using our result
$\rho_{00}=1/3$,
the distribution of $\theta$ is isotropic. In experiment one indeed finds that the
distribution is isotropic in $\theta$.
The experimental results at $\sqrt s =M_Z$ are:
\begin{eqnarray}
 \rho_{00} &=& 0.32\pm 0.04\pm0.03,\ \ \ \ {\rm
 DELPHI\cite{DB}} \nonumber \\
 \rho_{00}&=&0.33\pm0.06\pm0.05,\ \ \ \
 {\rm ALEPH\cite{AB}}, \nonumber \\
 \rho_{00}&=&0.36\pm0.06\pm0.07,\ \ \ \
 {\rm OPAL\cite{OB}}.
\end{eqnarray}
These results agree well with our prediction $\rho_{00}=1/3$.
It is interesting to note that the spin alignment is also
studied with the LUND string fragmentation model\cite{LUND} implemented
with JETSET\cite{JET} in \cite{XLL}, and the result $\rho_{00} = 0.567$
is obtained, which is not in agreement with experiment. In \cite{FP}
the $B^*$- and $B$ meson is taken as a bound state of  a $b$-quark and other light
degrees of freedom, and these light degrees have the total angular momentum
$j=\frac{1}{2}$. Because the parity is conserved in the formation of $B^*$,
the light degrees are produced with equal probability for $j_3=\pm\frac{1}{2}$.
With this argument one obtains the probability for a left-handed $b$-quark
to form a $B^*$- and $B$ meson is:
\begin{equation}
  P(\bar B^*(\lambda=-1)):P(\bar B^*(\lambda=0)):
  P(\bar B^*(\lambda=1)):P(\bar B) =\frac{1}{2} : \frac{1}{4} :0:\frac{1}{4} .
\end{equation}
One then also obtains $\rho_{00}=1/3$ in agreement with experiment and with our
results. One can also obtains the ratio of the production rates
\begin{equation}
  P_V = \frac {\sigma (B^*)}{\sigma (B^*) +\sigma (B)} =\frac{3}{4},
\end{equation}
this prediction is well tested with experiment:
\begin{eqnarray}
 P_V &=& 0.72\pm 0.03\pm0.06,\ \ \ \ \ \ {\rm
 DELPHI\cite{DB}} \nonumber \\
 P_V &=& 0.771\pm0.026\pm0.07,\ \ \
 {\rm ALEPH\cite{AB}}.
\end{eqnarray}
However the prediction of the ratio for c-flavored mesons  has a deviation
at $20\%$ level. It should be noted that in \cite{FP} only the diagonal
part of the density matrix is predicted, while in our work the predictions
are given for the complete matrix.
\par
The polarization measurement for $D^*$-meson
has been done with different $\sqrt s$, in some
experiments the non-diagonal part of the spin density matrix has also been
measured by measuring azimuthal angular distribution in $D^*$ decay,
where the decay mode into two pseudo-scalars, i.e.,
$D^*\to D\pi$, is used. Denoting
$\theta$ as the angle between the moving directions
of $D^*$ and of $\pi$ in the $D^*$-rest frame and $\phi$
as the azimuthal angle of $\pi$, then the angular distribution of $\pi$
is given by $W_{D^*\to D\pi}(\theta,\phi)\propto \sum_{\lambda\lambda'}
\rho_{\lambda\lambda'}Y_{1\lambda}(\theta,\phi)
Y^*_{1\lambda'}(\theta,\phi)$. Integrating over $\phi$ and using our result
$\rho_{00}=1/3$,
the distribution of $\theta$ is again isotropic. The experimental results
are summarized in Table 1 and also partly summarized in \cite{THK}.
\par\vfil\eject
\par\vskip20pt
\begin{center}
\centerline{{\bf Table 1}. Experimental Results for $D^*$}
\vspace{15pt}
\begin{tabular}{lll}
\hline\hline
  Collaboration\hspace{1cm} & $\sqrt s $\ in GeV\hspace{3cm} &  Results \\
  \hline
  CLEO\cite{CLEO} & 10.5  & $\rho_{00}=0.327\pm0.006$ \\
  HRS\cite{HRS} & 29    & $\rho_{00}=0.371\pm0.016$\\
  \ \           & \ \   & $\rho_{1-1}=0.04\pm0.03$ \\
  \ \           & \ \   & $\rho_{10}=0.00\pm0.01$ \\
  TPC\cite{TPC} & 29    & $\rho_{00}=0.301\pm0.042\pm0.007$\\
  \ \           &       & $\rho_{1-1}=0.01\pm0.03\pm0.00$ \\
  \ \           &       & $\rho_{10}=0.03\pm0.03\pm0.00$\\
  SLD\cite{SLD} & 91    & $\rho_{00}=0.34\pm0.08\pm0.13$\\
  OPAL\cite{OB} & 91    & $\rho_{00}=0.40\pm0.02\pm0.01$\\
  \ \           &  \ \  & $\rho_{1-1}=-0.039\pm0.014$\\
  \hline
\end{tabular}
\end{center}
\par\vskip15pt
From Table 1. we can see that the $\rho_{00}$ measured
by all experimental groups
is close to the prediction $\rho_{00}=\frac{1}{3}$,
the most precise result is obtained by CLEO, its
deviation from the prediction is $2\%$, the largest
deviation of the prediction is from the result made by OPAL
at $\sqrt s =90$GeV, it is $20\%$. In general, $\rho_{00}$
depends on the energy of $H^*$. Our results in Eq.(17) give
that $\rho_{00}$ is a constant in the heavy quark limit,
or the energy dependence is suppressed by $m_Q^{-2}$.
In experiment only a very weak energy dependence
is observed, e.g., in CLEO results\cite{CLEO}.
From our results $\rho_{1-1}$
is exactly zero in the heavy quark limit, the results from
TPC and from HRS  are in consistent with our result, a
non zero value is obtained by OPAL, which has a $3\sigma$
deviation from zero. These deviations may be explained
with effects of higher orders in $m_c^{-1}$, these
effects are expected to be substantial, because
$m_c$ is not so large. It is interesting to note only
results from OPAL at $\sqrt s =91$GeV have the largest
deviations from our predictions, while results from other
groups agree well with our predictions. At
$\sqrt s =10.5{\rm GeV\ or\ } 29$GeV, the effect of
the $Z$-boson exchange can be neglected, hence
the parity is conserved. We obtain $\rho_{10}=0$.
This prediction is also in agreement with the experimental result
made by TPC and by HRS.
\par
Since our results are derived without knowing the total
angular momentum $j$ of the light degrees of freedom in the heavy
meson, the agreement of our results with experiment
can not be used to extract the information
about $j$ from the experimental data in Eq.(19) and in Table 1.,
although $\rho_{00}=1/3$ can also be obtained by taking $j=1/2$.
One way to extract $j$ may be to measure the difference between
$\rho_{11}-\rho_{-1-1}$, but it seems not
possible, because the polarization
of $H^*$ is measured through its parity-conserved decay and
the polarization of decay products is not observed in experiment.
In the heavy quark limit, the nondiagonal element $\rho_{1-1}$
and $\rho_{-11}$ are zero, while the other nondiagonal
matrix elements are nonzero if the parity is not conserved
and the initial state is unpolarized. At higher orders in $m_Q^{-1}$
this can be changed, e.g., $H^*$ can have tensor polarization.
\par\vskip20pt\noindent
{\bf 3. The spin density matrix for spin-2 heavy meson}
\par
\vskip20pt
In this section we consider a spin-2 heavy meson $\H2$, which
contains one heavy quark $Q$. We study the inclusive production
\begin{equation}
e^+(\bp) +e^-(-\bp) \to \H2(\bk)+X.
\end{equation}
With the detailed analysis in the last section
it is straightforward to obtain the spin density
matrix. To avoid introducing too many notations, we
will use some notations used in the last section.
The used notations in this section are referred
to $\H2$. We denote the unnormalized spin density matrix
$R(\lambda,\lambda',\bp,\bk)$, which is defined by replacing
$H^*$ with $\H2$ in Eq.(2). The helicity $\lambda$ takes the
value from -2 to 2. We work in the $\H2$-rest frame, which is
related to the moving frame only through a Lorentz boost.
The $z$-direction is chosen as the moving direction of $\H2$.
In the rest frame the polarization tensor
of $\H2$ is $\epsilon_{ij}(\lambda)$, it has the following
properties:
\begin{eqnarray}
\epsilon_{ij}(\lambda)&=&\epsilon_{ji}(\lambda),\ \ \
\epsilon_{ii}(\lambda)=0, \nonumber\\
\sum_\lambda \epsilon_{ij}(\lambda)\epsilon_{kl}(\lambda)
 &=& \frac{1}{2}(\delta_{ik}\delta_{jl}+\delta_{il}\delta_{jk})
    -\frac{1}{3} \delta_{ij}\delta_{kl}.
\end{eqnarray}
We introduce a creation operator for $\H2$ in its rest-frame:
\begin{equation}
 \vert \H2(\lambda) \rangle = \epsilon_{ij}(\lambda) a^\dagger_{ij}
   \vert 0\rangle,
\end{equation}
where $a^\dagger_{ij}$ is symmetric and trace-less.
In analogy to Eq.(12) for the spin-1 case two operators
can be defined:
\begin{equation}
O(\H2) = \frac{1}{6} {\rm Tr} \psi a^\dagger_{ij}
      a_{ij} \psi^\dagger, \ \ \ \
O_s(\H2)=\frac{i}{12} {\rm Tr} \sigma_i \psi a^\dagger_{jl}
   a_{lk} \psi^\dagger \varepsilon_{ijk}.
\end{equation}
With these operators the matrix for $\H2$ corresponding
to that for $H^*$ in Eq. (13) can be written:
\begin{eqnarray}
\sum_X \langle 0\vert \psi \vert \H2(\lambda')X\rangle
   \langle X\H2(\lambda)\vert \psi^\dagger \vert 0\rangle
 &=& \frac{1}{5}\langle 0 \vert O(\H2) \vert 0 \rangle
       \epsilon^*_{ij}(\lambda)\epsilon_{ij}(\lambda') \cdot I
       \nonumber\\
  & & + \frac{i}{5}\langle 0 \vert O_s(\H2) \vert 0 \rangle
     \epsilon_{jk}^*(\lambda)\epsilon_{kl}(\lambda')
     \varepsilon_{ijl} \cdot \sigma_i,
\end{eqnarray}
the matrix is diagonal in color space. Predictions can be obtained
by replacing the matrix element in r.h.s. of Eq.(11) with that given above
and perform the same calculation as that in the last section. The unnormalized
spin density matrix for $\H2$ then reads:
\begin{equation}
R(\lambda,\lambda',\bp,\bk) = \frac{1}{5} a(\bp,\bk)
 \langle 0 \vert O(\H2) \vert 0 \rangle
       \epsilon^*_{ij}(\lambda)\epsilon_{ij}(\lambda')
    +\frac{i}{5} b_i(\bp,\bk) \langle 0 \vert O_s(\H2) \vert 0 \rangle
     \epsilon_{jk}^*(\lambda)\epsilon_{kl}(\lambda')
     \varepsilon_{ijl},
\end{equation}
where the coefficients $a(\bp,\bk)$ and $b_i(\bp,\bk)$ for $i=1,2,3$ are
the same as those defined in Eq.(16). With this matrix we obtain the normalized
spin density matrix for $\H2$:
\begin{equation}
\rho({\bf p},{\bf k}) =\frac{1}{5}\left(
\begin{array}{ccccc}
  1+P_3, & -\frac{1}{\sqrt 2}P_+, & 0,& 0, & 0 \\
   -\frac{1}{\sqrt 2}P_-, & 1+\frac{1}{2}P_3, & -\frac{\sqrt 3}{2}P_+, & 0, & 0 \\
  0, &-\frac{\sqrt 3}{2}P_-,  & 1, & -\frac{\sqrt 3}{2}P_+, & 0 \\
  0, & 0, & -\frac{\sqrt 3}{2}P_-, & 1-\frac{1}{2}P_3, & -\frac{1}{\sqrt 2}P_+ \\
  0, & 0, & 0, & -\frac{1}{\sqrt 2}P_-, & 1-P_3
\end{array}\right),
\end{equation}
with
\begin{equation}
P_3= \frac{b_3(\bp,\bk)}{a(\bp,\bk)}
            \cdot \frac{\langle 0 \vert O_s(\H2) \vert 0 \rangle}
             {\langle 0 \vert O(\H2) \vert 0 \rangle},\ \ \
P_\pm = \frac{b_1(\bp,\bk)\pm ib_2(\bp,\bk)}{\sqrt{2} a(\bp,\bk)}
           \cdot \frac{\langle 0 \vert O_s(\H2) \vert 0 \rangle}
             {\langle 0 \vert O(\H2) \vert 0 \rangle},\ \ \
\end{equation}
The indices of matrices in Eq.(29) run from -2 to 2. The results are  similar
to those for the spin-1 case.
The correction to the predictions
in Eq.(29) is of order of $m_Q^{-2}$ or of higher orders. The same comments
made in the last section for spin-1 heavy meson also apply here.
\par
It would be interesting to compare the predictions with experiment.
Unfortunately, there is no experimental data for $B^{**}$ meson,
there is a candidate $B_J^*(5732)$ observed in experiment, but
its spin is not determined. For c-flavored meson, a spin-2 meson
noted as $D_2^*(2460)$ is observed, and the information of its spin
is analyzed by ARGUS\cite{ARGUS} with a limited data sample. Through
the angular distribution of the decay $D_2^* \to D\pi$, it is
found that there is no significant population of
$\vert \lambda\vert =2$ states.
For the decay there is one form-factor in the decay amplitude, the angular
distribution can then be written as:
\begin{equation}
\frac{d\Gamma}{\Gamma d\phi d\cos\theta }(D_2^*\to D\pi)
 \sim \sum_{\lambda\lambda'} \rho_{\lambda\lambda'}
   Y_{2\lambda}(\theta,\phi) Y_{2\lambda'}^*(\theta,\phi),
\end{equation}
where $\theta$ is the angle between the moving directions
of $D_2^*$ and of $\pi$ in the $D_2^*$-rest frame, $\phi$
is the azimuthal angle of $\pi$. With our results we obtain:
\begin{equation}
\frac{d\Gamma}{\Gamma  d\cos\theta }(D_2^*\to D\pi)
 =\frac{1}{2},
\end{equation}
i.e., the distribution is isotropic. In the approach
in \cite{FP} the distribution is predicted as:
\begin{equation}
  \frac{d\Gamma}{\Gamma d\cos\theta }(D_2^*\to D\pi) =
 \frac{1}{4} [ 1+3\cos^2\theta -6\omega_{3/2}
    (3\cos^2\theta-1)],
\end{equation}
where an unknown parameter $\omega_{3/2}$ is introduced,
this parameter is interpreted as the probability to produce
a system of light degrees of freedom with the
total angular momentum $j=3/2$ and $j_3=3/2$,
and this system is combined with a heavy quark $Q$
to form $\H2$.
ARGUS has found that the data favored a small
$\omega_{3/2}$, indicating that a small portion of
$\vert \lambda\vert =2$ states is produced.
A re-analysis of the data is given in
\cite{FP}. Assuming
$\omega_{3/2}>0$, it is found\cite{FP}:
\begin{equation}
 \omega_{3/2}<0.24,\ \ \ 90\%\ {\rm C.L.}
\end{equation}
A model estimation
also gives a small $\omega_{3/2}$\cite{CM}.
With our prediction we obtain that
the distribution is isotropic.
This leads to $\omega_{3/2}=1/2$. It seems that experimental
results are against our predictions. However, one should
note that the correction from higher orders of $m_c^{-1}$
can be more significant than that in the case of $D^*$,
because the excited state $D_2^*$ is heavier than
$D^*$. The effects of higher orders may be estimated
by the mass difference:
\begin{equation}
\frac{ M(D_2^*)-M(D_0)}{M(D_0)} \approx 0.34,
\end{equation}
this indicates that the large correction of our
prediction $\omega_{3/2}=1/2$ exists. An extension
of our present work to the next-to-leading order
in $m_c^{-1}$ is needed to analyze this effect,
including effects from higher orders the distribution
will be not isotropic. At leading order of the expansion
in $m_Q^{-1}$ we obtain the relation
$\rho_{\lambda\lambda}+\rho_{-\lambda-\lambda}
=2/5$, this leads to that the distribution in Eq.(32) or (33) is
isotropic. If we add corrections at higher orders of $m_Q^{-1}$, the
relation will be changed, the coefficient $2/5$ receives
corrections which depend on $\lambda$. This will change the
distribution as predicted in Eq.(32) and it becomes non-isotropic.
It needs to be study further.
For $B^{**}$ meson the correction is expected to be small
because $m_b$ is large, hence the predictions
are more accurate.
It would be interesting to test our prediction
in future experiment with $B^{**}$.
\par\vskip20pt\noindent
{\bf 4. Summary}
\par\vskip15pt
In this work we have studied the polarization of a heavy meson in its
inclusive production, where the heavy meson contains one heavy quark.
A QCD factorization is performed for the spin density matrix
of a spin-1-and spin-2 meson. In the heavy quark limit,
the spin density matrix takes a factorized form, in which
the nonperturbative effect related to
the heavy meson is represented by two matrix elements defined
in HQET, while the perturbative part is related to the spin density
matrix of the heavy quark in its inclusive production. Hence, a
detailed relation between the spin density matrices of the heavy meson
and of the heavy quark is established.
\par
Without the knowledge of the
two matrix elements defined in HQET, some elements of the spin density
matrix can already be predicted. These elements are well measured
in experiment. We find that the measured element for $B^*$-meson at
LEP with $\sqrt s=91$GeV agrees well with our prediction. For
$D^*$-meson the polarization is measured at $e^+e^-$ colliders
with $\sqrt s =10.5,\ 29$ and 91GeV, the experimental results also agree
with our predictions, except a small deviation with OPAL data, the deviation
may be due to corrections from higher orders in $m_c^{-1}$. In general one
expects these corrections are large because $m_c$ is not so large,
the agreement for $D^*$-meson indicates that HQET can be well used
to predict the polarization of $D^*$-meson. For $D^{**}$ meson
not many experimental data are available. The results of a measurement
with large errors does not agree with our result. The disagreement
may be explained with large corrections from higher orders in $m_c^{-1}$,
these large corrections for $D^{**}$ meson are  more likely than
those for $D^*$ meson, because the binding energy of $D^{**}$ is larger
than that of $D^*$ meson.
\par
The polarization of a heavy meson in its inclusive production has also
been studied in \cite{FP}, where one takes the total angular momentum
of light degrees of freedom in the heavy meson to be the lowest of
the possible values. With this the diagonal part of the spin density matrix
can be predicted. In our work we only use the QCD factorization concept
by employing the expansion in the inverse of the heavy quark mass,
and we obtain results not only for the diagonal part, but also
for the nondiagonal part of the spin density matrix. For the spin-1
case we obtain the same matrix element $\rho_{00}$, while
for the spin-2 case, our results are different than those in
\cite{FP}, it leads to that the prediction for the angular
distribution of the decay $D^{**}\to D \pi$ is different than that
predicted in \cite{FP}. However, large corrections to our prediction
are likely possible, it will be interesting to test predictions
for $B^{**}$ meson in future experiments, because corrections
in the case with $B^{**}$ are expected to be small.
\par
Although we have given in this work detailed predictions
for inclusive production of a polarized heavy meson at
a $e^+e^-$ collider, our approach can be easily
generalized to other inclusive productions, testable predictions
can be made without a detailed calculation, for example, in inclusive
production of $B^*$ at an electron-hadron- or a hadron-hadron collider
we always have the prediction
$\rho_{00}=1/3$ and $\rho_{-11}=\rho_{1-1}=0$ in the heavy quark limit.

\par\vskip20pt

\vskip 10mm
\begin{center}
{\bf\large Acknowledgments}
\end{center}

The work is supported  by National Nature Science Foundation of P. R.
China and by the Hundred Young Scientist Program of Academia
Sinica of P. R. China.


\begin{thebibliography}{99}

\bibitem{HQET}  N. Isgur and M.B. Wise, Phys. Lett. B232 (1989) 113, ibid
B237 (1990) 527

E. Eichten and B. Hill, Phys. Lett. B234 (1990) 511

B. Grinstein, Nucl. Phys. B339 (1990) 253

H. Georgi, Phys. Lett. B240 (1990) 447

\bibitem{Review} M. Neubert, Phys. Rept. C245 (1994) 259

M. Shifman, Lectures given at Theoretical Advanced Study Institue {\em
QCD and Beyond}, Unversity Colorado, June 1975,  hep-ph/9510377

\bibitem{Ma} J.P. Ma, Nucl. Phys. B506 (1997) 329

\bibitem{FP} A.F. Falk and M.E. Peskin, Phys. Rev. D49 (1994) 3320

\bibitem{BeMa} W. Bernreuther, O. Nachtmann, P. Overmann and T. Schr\"oder Nucl. Phys. B388
    (1992) 53, Erratum-ibid. B406 (1993) 516

W. Bernreuther, J.P. Ma and T. Schroder, Phys.Lett. B297 (1992) 318

\bibitem{Ji} Xiangdong Ji, Phys. Rev. D49 (1994) 114

\bibitem{SST} A. Sch\"afer, L. Szymanowski and O.V. Teryaev,
Phys. Lett. B464 (1999) 94

\bibitem{DB} P. Abreu et al., DELPHI Collaboration,  Z. Phys. C68 (1995) 353

\bibitem{AB} D. Buskulic et al., ALEPH Collaboration, Z. Phys. C69 (1993) 393

\bibitem{OB} K. Ackerstaff et al., OPAL Collaboration, Z. Phys. C74 (1997) 437

\bibitem{LUND} B. Andersson et al., Phys. Rep. C97 (1983) 31

\bibitem{JET} T. Sj\"ostrand, Comput. Phys. Commun. 39 (1986) 347

\bibitem{XLL} Xu Qing-hua, Liu Chun-xiu and Liang Zuo-tang, Phys. Rev. D63 (2001) 111301(R)

\bibitem{THK} T.H. Kress, hep-ex/0101047, to be published in the proceedings
of 30th International Symposium on Multiparticle Dynamics (ISMD
2000)

\bibitem{CLEO} G. Brandenburg et al., CLEO Collaboration, Phys. Rev. D58 (1998) 052002

\bibitem{HRS} S. Abachi et al., HRS Collaboration, Phys. Lett. B199 (1987) 585

\bibitem{TPC} S. Aihara et al., TPC Collaboration, Phys. Rev. D43 (1991) 29

\bibitem{SLD} K. Abe et al., SLD Collaboration, presented at International Europhysics
Conference on High-Energy Physics (HEP97), Jerusalem, Isral, 1997, Report No.
SLAC-PUB 7574

\bibitem{ARGUS} H. Albrecht et al., ARGUS Collaboration, Phys. Lett. B221 (1989) 422,
Phys. Lett. b232 (1989) 398

\bibitem{CM}  Yu-Qi Chen, M.B. Wise, Phys.Rev. D50 (1994) 4706

\end{thebibliography}
\end{document}